\documentclass[10pt,journal,twocolumn]{IEEEtran}
\usepackage{amsmath,amssymb,cite,url,array,booktabs,xcolor,graphicx,svg,tabularx,placeins}

\AtBeginDocument{\lefthyphenmin=3\righthyphenmin=3}

\newif\ifshoweditnotes
\showeditnotesfalse
\definecolor{editcolor}{RGB}{150,38,38}
\definecolor{boxfill}{RGB}{246,248,250}
\newcommand{\figplaceholder}[2]{%
  \begingroup\setlength{\fboxsep}{7pt}%
  \fcolorbox{black!45}{boxfill}{%
    \begin{minipage}{\dimexpr\columnwidth-2\fboxsep-2\fboxrule\relax}
      \small\textbf{#1}\par\smallskip #2
    \end{minipage}}%
  \endgroup}

\newcommand{\figureasset}[3]{%
  \IfFileExists{#1}{%
    \includegraphics[width=\columnwidth]{#1}%
  }{%
    \figplaceholder{#2}{#3}%
  }}

\newcolumntype{L}[1]{>{\raggedright\arraybackslash}p{#1}}
\DeclareMathOperator*{\argmin}{arg\,min}

\renewcommand{\arraystretch}{1.12}
\title{Agentic TCAD Calibration Workflow for Oxide Semiconductor Transistors}

\author{Gyujun~Jeong, Junmo~Lee, Sungwon~Cho, Woohyun~Hwang,
Kwangyou~Seo, Suhwan~Lim, Wanki~Kim, Daewon~Ha, Rishi~Ranade,
Kihang~Youn, Ram~Cherukuri, Yiyi~Wang, Asif~Khan, and~Shimeng~Yu%
\thanks{Gyujun Jeong, Junmo Lee, Sungwon Cho, Asif Khan, and Shimeng Yu are with the School of Electrical and Computer Engineering, Georgia Institute of Technology, Atlanta, GA, USA (e-mail: gjeong35@gatech.edu; shimeng.yu@ece.gatech.edu).}%
\thanks{Woohyun Hwang, Kwangyou Seo, Suhwan Lim, Wanki Kim, and Daewon Ha are with Semiconductor Research and Development, Samsung Electronics Co., Ltd., South Korea.}%
\thanks{Rishi Ranade, Kihang Youn, Ram Cherukuri, and Yiyi Wang are with NVIDIA, Santa Clara, CA, USA.}}
\begin{document}
\maketitle
\begin{abstract}
Experimental TCAD calibration is essential for predictive technology modeling of emerging oxide semiconductor transistors. However, it remains time-consuming and expert dependent because of model ambiguity. Multiple physical models and parameter sets can reproduce the same measured transfer characteristics, while local fitting alone cannot uniquely identify the underlying device physics. We present the first demonstration of an agentic TCAD calibration workflow for a fabricated bottom-gate In--W--O (BG-IWO) transistor. Starting from the measured transfer curve and device information, the workflow uses measurement--TCAD residuals and local sensitivity tests to select bounded parameter corrections or evaluate additional physical models, and accept only updates that improve device metrics. The LLM agent orchestrates the workflow, while Sentaurus governs the device physics. For the 2\%-W reference device, five agent-suggested updates yield a fixed calibrated model, reducing the multi-metric device objective $J$ by 14.3$\times$. Maximum $V_{\mathrm{th}}$/$I_{\mathrm{on}}$ errors are 36.1~mV/0.022 decade for varying-drain-bias tests and 46.2~mV/0.062 decade for varying-channel-length tests, demonstrating model transferability across bias and geometry rather than a local parameter fit. W-composition tests provide process-sensitive insight. This agentic workflow provides a faster route to model development for emerging device technologies.

\end{abstract}

\section{Introduction}
Oxide semiconductor transistors have emerged as BEOL-compatible, low-thermal-budget candidates for CMOS+X applications such as memory access transistors and power converters~\cite{ye2020edram,deng2023beolpower}. For emerging devices under short-loop experimentation, turning measured transfer characteristics into a predictive TCAD deck for future process optimization remains a nontrivial inverse device-modeling problem. In practice, calibration becomes a time-consuming and expert-dependent bottleneck because each iteration requires physical-model selection, parameter adjustment, deck implementation, Sentaurus execution, and interpretation of measurement--simulation residuals. The measured transfer characteristics reflect coupled electrostatic, transport, and trapping effects, while their TCAD reproduction also depends on device geometry, parameter values, meshing, and numerical implementation.

Fig.~1 frames this bottleneck as TCAD model-search ambiguity. Multiple physics models and parameter choices can fit the same measured transfer curve, while a local parameter fit alone does not identify the underlying device physics or guarantee model generalization. This work therefore reduces the active search effort by using measurement--TCAD residuals to rank TCAD physical models and parameter groups, run bounded Sentaurus tests, and accept only updates that improve predefined device metrics. 

\begin{figure}[!tbp]
\centering
\figureasset{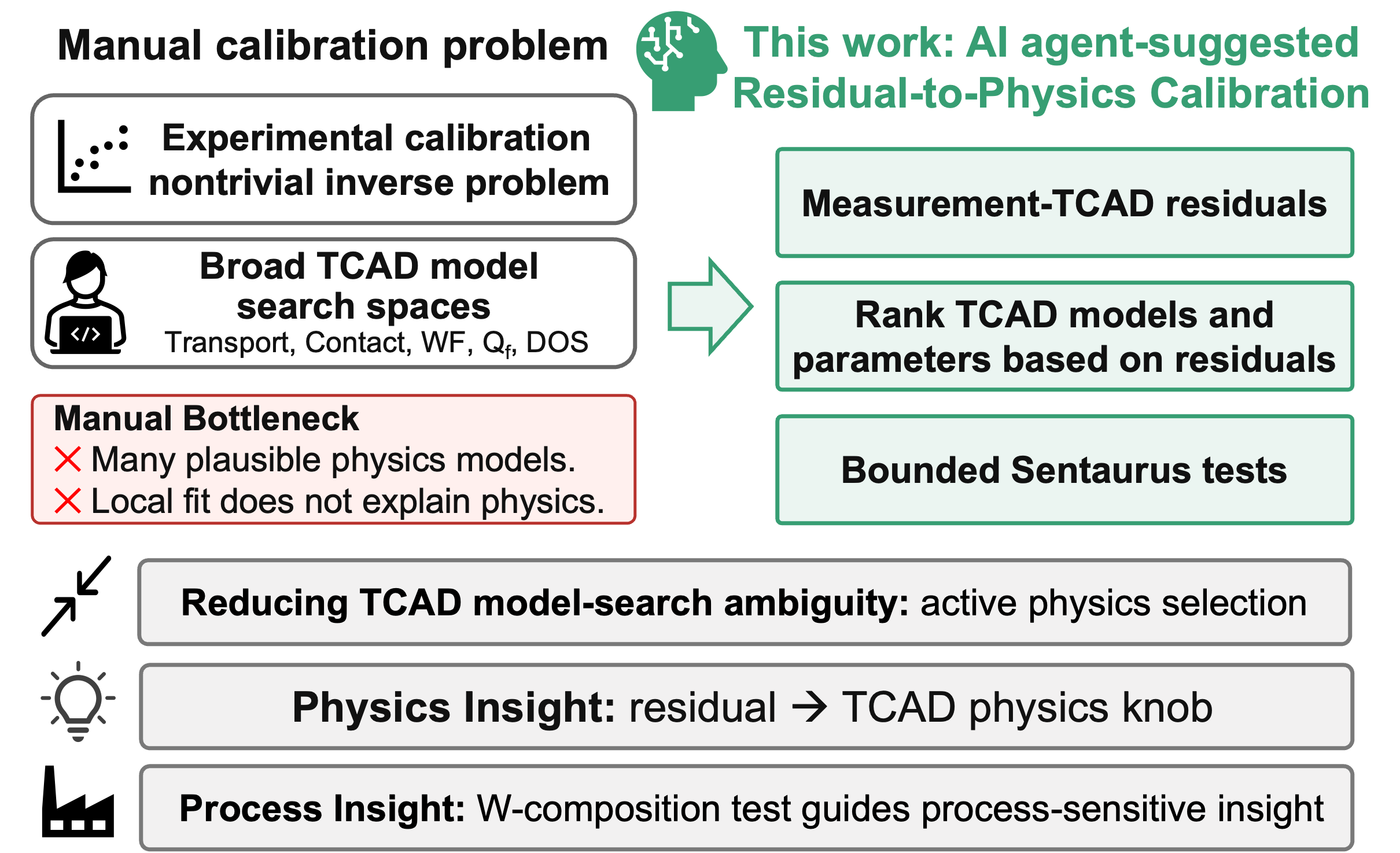}
{}{}
\caption{Motivation for agentic TCAD calibration. Manual calibration is a nontrivial inverse problem because multiple TCAD physics models can fit the same measured transfer curve. This work uses measurement-TCAD residuals to rank model classes, run Sentaurus, and accept only updates that improve predefined device metrics, orchestrated by an AI agent. The workflow produces a fixed calibrated model, reduces model/parameter ambiguity, and provides physical and process insight for follow-up TCAD studies.}
\label{fig:motivation}
\end{figure}

Prior machine learning (ML) and artificial intelligence (AI)-assisted TCAD and modeling studies have addressed adjacent tasks that differ from this work, including TCAD--ML variation and temperature analysis~\cite{wong2020tcadml}, inverse parameter calibration~\cite{ng2025tcadtft}, and large language model (LLM)-based agentic TCAD code generation and device optimization~\cite{agentictcad2026}. However, these earlier works operate within predefined physics, parameter, or design spaces. Hence, they do not provide physical insights for automating model selection and parameter calibration based on experimental data. Table~\ref{tab:priorwork} summarizes the distinctions from prior AI/ML-assisted TCAD studies.

\begin{table*}[!t]
\caption{Comparison with prior AI/ML-assisted TCAD studies.}
\label{tab:priorwork}
\centering
\footnotesize
\setlength{\tabcolsep}{3.5pt}
\renewcommand{\arraystretch}{1.15}
\begin{tabularx}{\textwidth}{@{}
  >{\raggedright\arraybackslash}p{0.12\textwidth}
  >{\raggedright\arraybackslash}X
  >{\raggedright\arraybackslash}X
  >{\raggedright\arraybackslash}X
  >{\raggedright\arraybackslash}p{0.23\textwidth}@{}}
\toprule
\textbf{Aspect} &
\textbf{TCAD--ML}~\cite{wong2020tcadml} &
\textbf{Inverse TCAD calibration}~\cite{ng2025tcadtft} &
\textbf{AgenticTCAD}~\cite{agentictcad2026} &
\textbf{This work} \\
\midrule

\textbf{Objective} &
Analyze device variation and temperature effects &
Infer parameters of a predefined TCAD model &
Generate TCAD code and optimize device designs &
Experimentally calibrate TCAD physics and parameters to measured $I$--$V$ \\

\textbf{Inputs} &
TCAD-generated $I$--$V$ database &
TCAD-generated training data and measured TFT $I$--$V$ &
Natural-language design specifications &
Measured $I$--$V$, device information, and TCAD residuals \\

\textbf{Search space} &
Fixed TCAD physics, sampled variation space &
Predefined physics model and fitted parameters &
Predefined design space &
Reduced search space by active physics model selection by reported physical insights \\

\textbf{Decision basis} &
ML surrogate prediction &
AI-based inverse fitting &
LLM agent--simulator iteration &
Agentic workflow with residual-guided parameter and model selection\\

\textbf{Validation} &
Experimental demonstration &
TFT calibration cases &
Simulation-defined target specifications &
Drain-bias and channel-length transfer tests without retuning \\

\bottomrule
\end{tabularx}
\end{table*}

To the best of our knowledge, this work presents the first experimental demonstration of an agentic TCAD calibration workflow. The workflow integrates (i) physics-model selection using retrieval-augmented generation (RAG)~\cite{lewis2020rag}, (ii) on-the-fly TCAD execution and evaluation of intermediate results, (iii) residual-guided selection of subsequent bounded calibration tests, and (iv) extraction of device-physics and process insights, orchestrated by the AI agent.

\section{Agentic TCAD Calibration Workflow}
\subsection{Workflow Overview and Responsibilities}
\begin{figure}[!tbp]
\centering
\figureasset{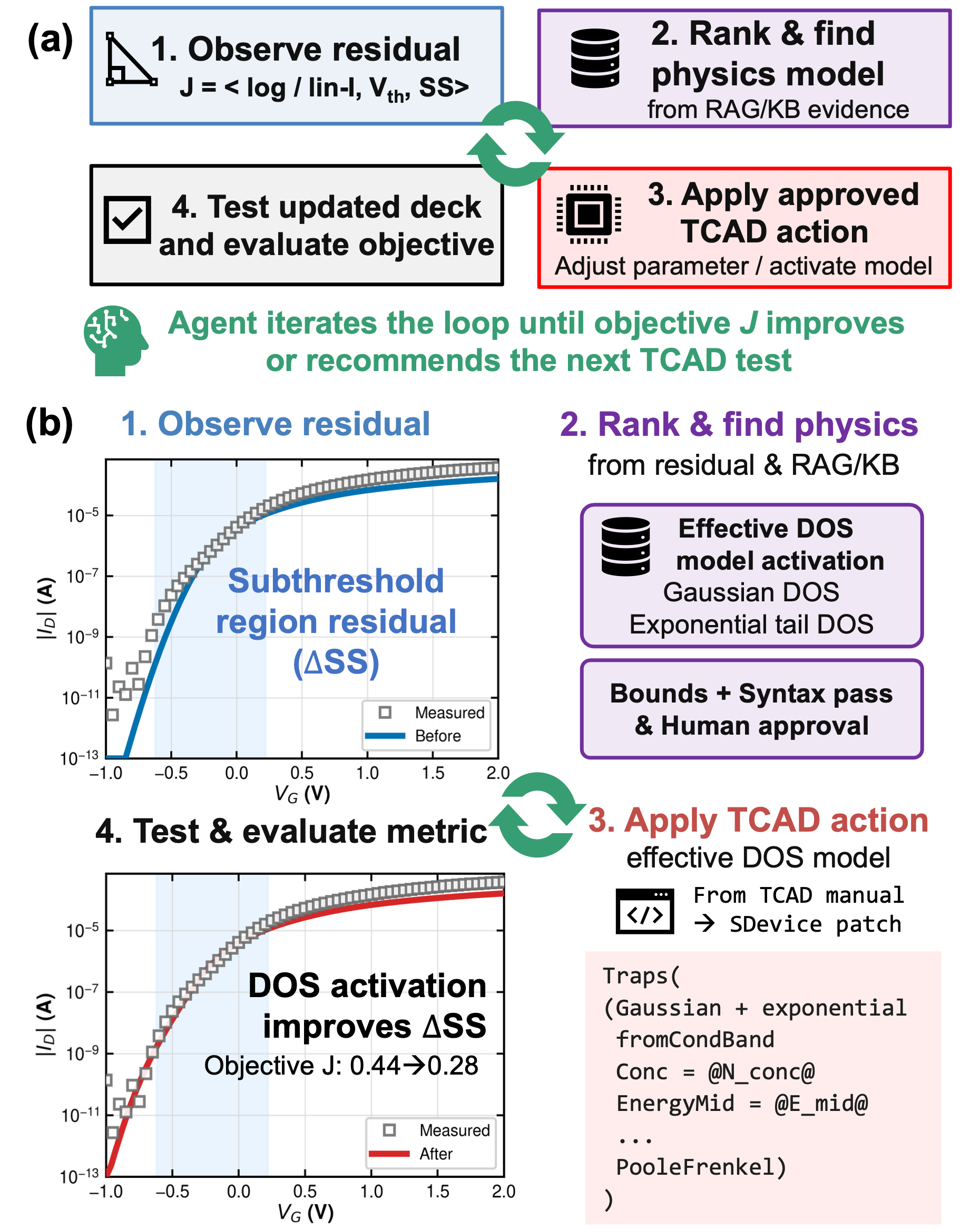}{}{}
\caption{Residual-guided TCAD calibration loop. (a) The agentic workflow demonstrates how residuals guide the next TCAD update. The agent compares the measured transfer curve with the current TCAD model, extracts residuals, and suggests either a parameter adjustment or model activation. (b) The example shows how a subthreshold swing residual ($\Delta$SS) drives Gaussian/tail DOS activation, improving the multi-metric device objective $J$.}
\label{fig:workflow}
\end{figure}

Fig.~\ref{fig:workflow}(a) shows the residual-guided TCAD calibration loop. The workflow compares the measured transfer curve with the current TCAD result, extracts the residuals and device metrics, and determines whether the next action should adjust an existing parameter or evaluate an additional physics model. RAG retrieves supporting evidence from a knowledge base (KB) comprising the Sentaurus Device User Guide~\cite{sentaurusdeviceguide} and reviewed device literature. The proposed parameter or physics-model update is translated into Sentaurus syntax, executed, and evaluated using the multi-metric device objective $J$ defined in~\ref{sec:parameter_adjustment}. Each update is accepted only when it reduces $J$. Otherwise, the update is rejected, and the workflow recommends the next bounded TCAD test.

Fig.~\ref{fig:workflow}(b) illustrates an example flow. The remaining mismatch is mainly in the subthreshold swing, as indicated by $\Delta\mathrm{SS}$. A RAG search of the Sentaurus manual and device literature identifies Gaussian and tail DOS models whose documented effects are consistent with the measured subthreshold shape. After human review and checks of the parameter ranges and SDevice syntax, these terms are added to the deck. The subsequent Sentaurus run reduces $J$. In the calibration loop, bounded one-factor-at-a-time (OFAT) probes identify existing parameter groups that affect the dominant residual. The selected residual vector then guides derivative-free Gauss--Newton (DFO-GN) updates~\cite{cartis2019dfogn}, which are suitable for expensive live-TCAD residual evaluations. Section~\ref{sec:parameter_adjustment} defines the local sensitivity, and the DFO-GN update. Section~\ref{sec:model_activation} describes when an additional physical model is considered and how each update is verified. 

Table~\ref{tab:agent_human_tasks} distinguishes the responsibilities of the LLM agent, Sentaurus TCAD, and the human reviewer. In an LLM-agent workflow, the language model can coordinate external tools and use the returned results to select subsequent actions~\cite{wang2024llmagents}. Here, the LLM agent serves as the workflow orchestrator rather than the device solver. Sentaurus TCAD solves the device equations and computes the device response for each proposed update. The human reviewer defines and approves the TCAD workflow instructions provided to the agent through \emph{Agent Skills}~\cite{nvidiaskills}, including permitted deck modifications, calibration-metric checks, and numerical-recovery procedures. The workflow logic is implemented independently of the selected LLM backend. The reported implementations used OpenAI GPT-5.5~\cite{openai2026gpt55} as the backend LLM and executed the workflow through OpenAI Codex~\cite{openai2026codex} and the open-source NVIDIA NeMo Agent Toolkit~\cite{nemoagenttoolkit}. All device simulations were performed using Synopsys Sentaurus W-2024.09-SP1.

\begin{table}[!t]
\caption{Responsibilities in the agentic TCAD workflow.}
\label{tab:agent_human_tasks}
\centering
\footnotesize
\setlength{\tabcolsep}{3.5pt}
\renewcommand{\arraystretch}{1.12}
\begin{tabularx}{\columnwidth}{@{}
  >{\raggedright\arraybackslash}p{0.28\columnwidth}
  >{\raggedright\arraybackslash}X@{}}
\toprule
\textbf{Component} & \textbf{Responsibility} \\
\midrule

\textbf{LLM agent} &
Analyze residuals and device metrics based on the retrieved RAG/KB evidence, and propose a bounded TCAD update. \\

\textbf{Sentaurus TCAD} &
Execute the human-approved deck and return the simulated transfer characteristics and convergence status. \\

\textbf{Human reviewer} &
Provide the device stack and test splits, approve agent-proposed updates, and verify their physical validity and interpretation. \\

\bottomrule
\end{tabularx}
\end{table}

\subsection{Device Structure and Experimental Design}
\begin{figure}[!tbp]
\centering
\figureasset{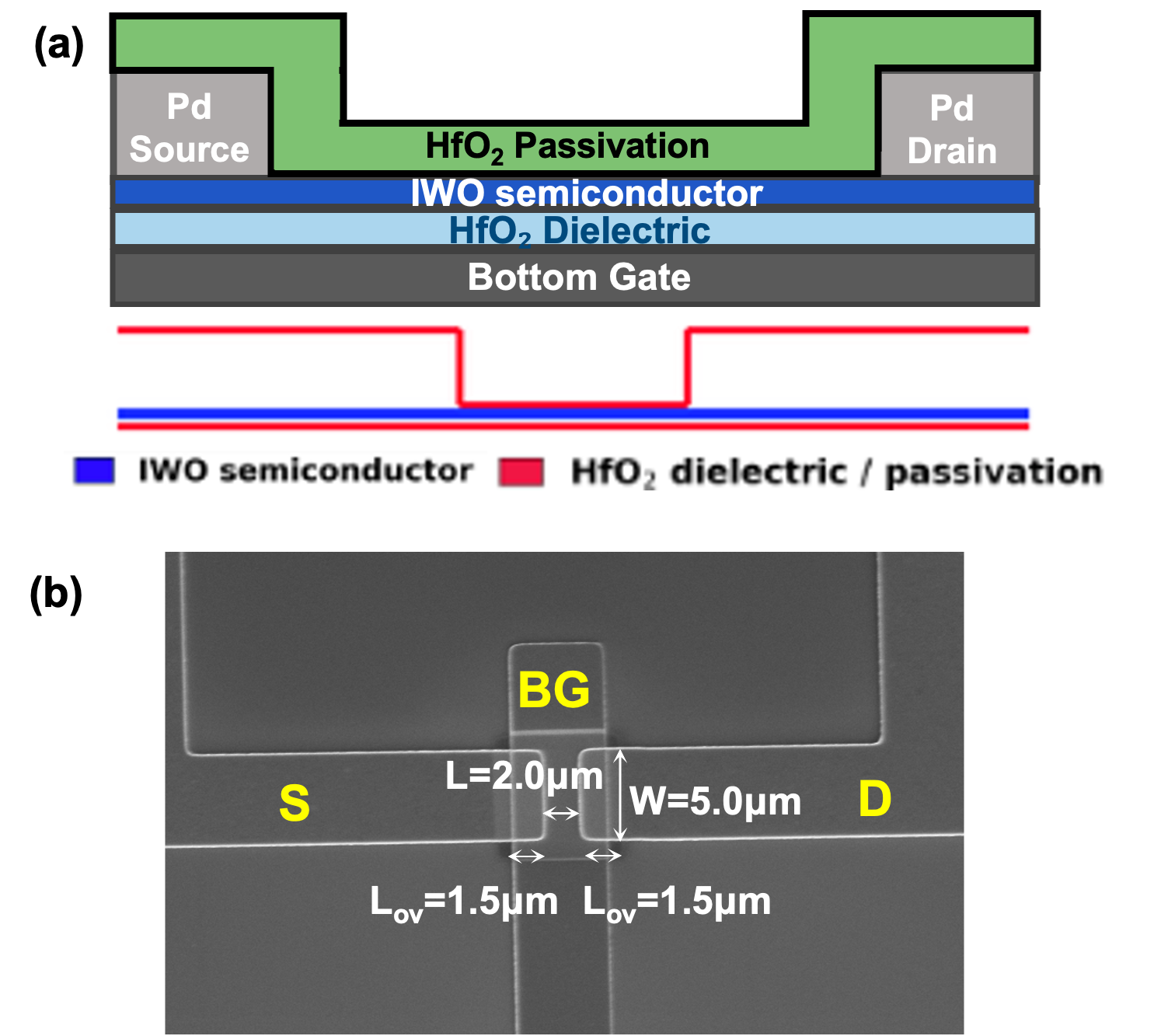}{}{}
\caption{(a) Fixed 2D TCAD geometry schematic and SVisual visualization of the BG-IWO. (b) SEM top view of an $L=2~\mu\mathrm{m}$ BG-IWO transistor used as a dimension reference.}
\label{fig:device}
\end{figure}

The experimental target is a fabricated bottom-gate In--W--O (BG-IWO) transistor measured using forward $I_D$--$V_G$ sweeps. Table~\ref{tab:device_specs} summarizes the device dimensions, materials, and measurement configuration. Fig.~\ref{fig:device}(a) shows the fixed 2D TCAD geometry and its SVisual visualization, and Fig.~\ref{fig:device}(b) provides SEM dimension references. Device geometry and basic material information are fixed before fitting, while the underlying physical model parameters are to be calibrated.

\begin{table}[!t]
\caption{Fabricated BG-IWO transistor and device specifications used for calibration.}
\label{tab:device_specs}
\centering
\footnotesize
\setlength{\tabcolsep}{3.5pt}
\renewcommand{\arraystretch}{1.12}
\begin{tabularx}{\columnwidth}{@{}
  >{\raggedright\arraybackslash}p{0.34\columnwidth}
  >{\raggedright\arraybackslash}X@{}}
\toprule
\textbf{Parameter} & \textbf{Specification} \\
\midrule
Measurement & Forward $I_D$--$V_G$ sweep \\
Channel material & Amorphous In--W--O (IWO): 2\% W (calibration) and 4\% W
(process test) \\
Channel length, $L$ & $1.0~\mu\mathrm{m}$ (calibration) and
$2.0~\mu\mathrm{m}$ (length test) \\
Channel width, $W$ & $5.0~\mu\mathrm{m}$ \\
S/D overlap, $L_{\mathrm{ov}}$ & $1.5~\mu\mathrm{m}$ per side
($3.0~\mu\mathrm{m}$ total) \\
IWO thickness & $3.5~\mathrm{nm}$ \\
Gate dielectric & HfO$_2$, $7.0~\mathrm{nm}$ \\
Passivation & Conformal HfO$_2$, $5.0~\mathrm{nm}$ \\
S/D contacts & Pd, $40~\mathrm{nm}$ \\
\bottomrule
\end{tabularx}
\end{table}

\begin{figure}[!t] \centering \figureasset{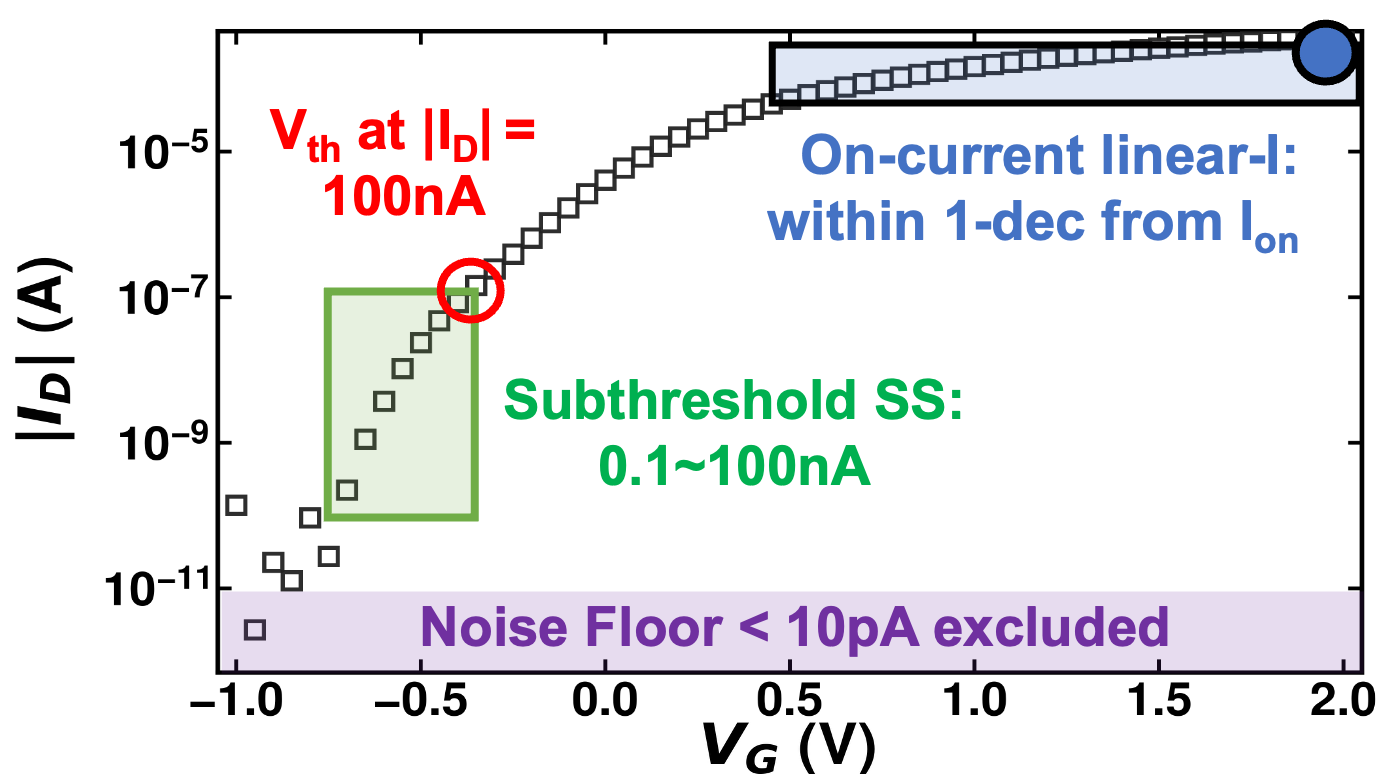}{}{} \caption{Calibration metric windows for objective $J$: log/linear-current, $V_{\rm th}$, $I_{\rm on}$, and SS residuals define the calibration objective in Eq.~\eqref{eq:calibrationJ}.} \label{fig:metric_windows} \end{figure}

As specified in Table~\ref{tab:data_splits}, nominal calibration uses the two measured $I_D$--$V_G$ transfer curves acquired at $V_D\in\mathcal B_{\rm cal}$, where $\mathcal B_{\rm cal}=\{50~\mathrm{mV},\,1~\mathrm{V}\}$. Fig.~\ref{fig:metric_windows} shows the metric windows used to evaluate these errors and extract $V_{\rm th}$, $I_{\rm on}$, and SS. The same current floor, gate-voltage windows, and metric-extraction method are applied at every parameter setting. Only the nominal 2\%-W, $L=1~\mu\mathrm{m}$ curves measured at biases $b\in\mathcal B_{\rm cal}$ contribute to $J$. All other drain-bias and channel-length conditions are reserved for evaluating the final calibrated model without retuning.

\begin{table}[!t]
\caption{Measurement data split.}
\label{tab:data_splits}
\centering
\footnotesize
\setlength{\tabcolsep}{3.5pt}
\renewcommand{\arraystretch}{1.12}
\begin{tabularx}{\columnwidth}{@{}
  >{\raggedright\arraybackslash}p{0.29\columnwidth}
  >{\raggedright\arraybackslash}X@{}}
\toprule
\textbf{Split} & \textbf{Conditions and model use} \\
\midrule
Nominal calibration &
2\% W, $L=1.0~\mu\mathrm{m}$, $V_D\in\mathcal B_{\rm cal}$, where $\mathcal B_{\rm cal}=\{0.05,1.00\}~\mathrm{V}$. Used for nominal calibration objective $J$ \\

Drain-bias transfer &
2\% W, $L=1.0~\mu\mathrm{m}$,
$V_D=\{0.05,0.10,0.20,0.50,1.00\}~\mathrm{V}$ for varying-bias-test without retuning\\

Channel-length transfer &
2\% W, $L=2.0~\mu\mathrm{m}$,
$V_D=\{0.05,1.00\}~\mathrm{V}$ for varying-channel-length test without retuning \\

W-composition test &
4\% W, $L=1.0~\mu\mathrm{m}$,
$V_D=\{0.05,1.00\}~\mathrm{V}$. Separately recalibrated for process sensitivity test \\
\bottomrule
\end{tabularx}
\end{table}

\subsection{Calibration objective J and sensitivity-based parameter adjustment}
\label{sec:parameter_adjustment}

For the drain-bias set $\mathcal B_{\rm cal}$ defined in Table~\ref{tab:data_splits}, the multi-metric calibration objective for a TCAD model $M$ with parameter vector $\boldsymbol{\theta}$ is \begin{equation} \begin{split} J(M,\boldsymbol{\theta}) = \frac{1}{|\mathcal B_{\rm cal}|} \sum_{b\in\mathcal B_{\rm cal}} \big[& w_{\log}E_{\log,b} +w_{\rm lin}E_{{\rm lin},b}\\ &+w_{\rm th}|\Delta V_{{\rm th},b}| +w_{\rm on}|\Delta\log_{10}I_{{\rm on},b}|\\ &+w_{\rm SS}|\Delta{\rm SS}_b| \big]. \end{split} \label{eq:calibrationJ} \end{equation} 

Here, $b$ denotes the drain bias, and $E_{\log,b}$ and $E_{{\rm lin},b}$ are the root-mean-square errors (RMSEs)  between the simulated and measured $\log_{10}|I_D|$ and $|I_D|$, respectively. As illustrated in Fig.~\ref{fig:metric_windows}, $V_{\rm th}$ is extracted at $|I_D|=100~\mathrm{nA}$, SS is obtained from the average inverse slope between $0.1$ and $100~\mathrm{nA}$, and $I_{\rm on}$ is the maximum $|I_D|$ in the transfer curve. The linear-current error is evaluated within one decade of $I_{\rm on}$, while currents below the $10$ pA measurement floor are excluded from the log-current error. Each $\Delta$ term denotes the simulation--measurement difference in the corresponding extracted metric. The fixed weights in Eq.~\eqref{eq:calibrationJ} are $w_{\log}=0.25$, $w_{\rm lin}=0.10$, $w_{\rm th}=1.00$, $w_{\rm on}=0.20$, and $w_{\rm SS}=0.25$. The same extraction rules are applied to every measured and simulated curve, and each objective term is normalized and weighted using values fixed before calibration.

The normalized signed residual and local sensitivity matrix for parameter group $P$ are
\begin{equation}
r_i=\frac{s_i(M,\boldsymbol{\theta})-m_i}{a_i},
\qquad
S_P=\frac{\partial\mathbf r}{\partial\mathbf x_P},
\label{eq:residual}
\end{equation}
where $s_i$ and $m_i$ are corresponding simulated and measured current values or extracted device metrics, $a_i>0$ is a fixed normalization factor chosen so that $r_i$ is dimensionless, and $\mathbf{x}_P$ denotes the normalized TCAD parameter for group $P$.

The scalar objective $J$ quantifies the overall fit between the simulated and measured transfer characteristics. The magnitude, sign, and gate voltage ($V_G$) dependence of $\mathbf r$ distinguish among a $V_{\rm th}$ offset, a drain current ($I_D$) magnitude mismatch, and an SS mismatch. The sensitivity $S_P$ indicates which existing TCAD parameter group can correct the dominant residual. Within the approved parameter bounds, OFAT Sentaurus simulations estimate the local sensitivity matrix $S_P$ by varying one TCAD parameter at a time and evaluating the resulting finite-difference
change in the residual vector.

For a parameter group $P$ selected by the OFAT sensitivity ranking, DFO-GN
calculates the bounded parameter change
\begin{equation}
\mathbf{s}_P^\star =
\argmin_{\substack{
\boldsymbol{\ell}_P
\leq \mathbf{x}_P+\mathbf{s}
\leq \mathbf{u}_P\\
\|\mathbf{s}\|_2\leq\Delta_P}}
\frac{1}{2}
\left\|W^{1/2}
\left(\mathbf{r}+S_P\mathbf{s}\right)\right\|_2^2 .
\label{eq:dfogn}
\end{equation}

Here, $\boldsymbol{\ell}_P$ and $\mathbf{u}_P$ are the allowed parameter
limits, $\Delta_P$ limits the parameter change in one update,
and $W$ contains fixed residual weights. Only parameters in group $P$
are changed. The local least-squares problem calculates a trial parameter
setting. The proposed change is evaluated by a full Sentaurus run, and the resulting $J$ determines whether the fit improves.

\subsection{Physics-Model Activation and Update Verification}
\label{sec:model_activation}

An additional physics model is considered only when a systematic transfer-curve mismatch persists after bounded adjustment of the existing parameters. The RAG search retrieves candidate models from the Sentaurus manual and reviewed device literature whose documented behavior is consistent with the remaining mismatch. The human reviewer verifies the applicability of the new physics model and parameter range before approving any deck modification. A parameter or model update is accepted only when the Sentaurus run converges and reduces $J$. Nonconverged trials undergo a predefined numerical-recovery sequence, such as adjusting the bias-step size, iteration limit, or solver tolerance. If convergence cannot be recovered, the run is recorded as a numerical failure and excluded from model-update decisions.


\section{Experimental Calibration and Model Transfer Test}
\subsection{Residual-Guided Calibration of the Reference Device}
Fig.~5 summarizes how five accepted updates yield a calibrated BG-IWO TCAD model. All TCAD results in this section use the fixed 2D structure defined in Table~\ref{tab:device_specs}. The agent suggests each tuning step by evaluating the residual components after each Sentaurus run. The baseline has a multi-metric objective of $J=2.589$. 

In Step~1, the agent identifies a low-current residual and suggests lowering the source/drain effective WF from 4.7 to 4.16~eV, which reduces $J$ to 1.537. Because the dominant residual in Step~2 is a threshold offset, the agent suggests lowering the gate WF from 4.4 to 4.0~eV and enabling a gate-interface fixed-charge term $Q_f=+1.59\times10^{10}\,\mathrm{cm}^{-2}$, reducing $J$ to 0.574. In Step~3, the residual indicates a charge deficit, leading the agent to increase $N_{D,\mathrm{eff}}$ from $1.0\times10^{18}$ to $2.56\times10^{18}\,\mathrm{cm}^{-3}$, reducing $J$ to 0.443. The left electrostatic/charge panel in Fig.~5(c) shows that the WF, \(Q_f\), and \(N_{D,\mathrm{eff}}\) updates first correct the electrostatic/charge response. 

After these updates, the remaining error is mainly a SS residual, as shown in Fig.~5(b) and Fig.~5(e). Step~4 therefore introduces the Gaussian and tail effective DOS descriptors shown in the right DOS panel of Fig.~5(c), reducing \(J\) to 0.282. The Gaussian term has \(N_{\mathrm{DOS}}=2.19\times10^{19}\,\mathrm{cm}^{-3}eV^{-1}\), centered at \(E_C-0.12\)~eV, and \(\sigma=0.08\)~eV; the tail term has \(N_{\mathrm{tail}}=7.8\times10^{19}\,\mathrm{cm}^{-3}eV^{-1}\), \(E_{\mathrm{tail}}=0.0\)~eV, and \(\sigma_{\mathrm{tail}}=0.133\)~eV. In Step~5, the agent suggests a final transport adjustment, changing $\mu_0$ from 10 to $24\,\mathrm{cm}^2/\mathrm{Vs}$, yielding the final calibrated model with $J=0.181$. The fitted variables are calibration-effective TCAD descriptors---effective work-function (WF) terms, gate-interface fixed charge ($Q_f$), effective donor-like background density ($N_{D,\mathrm{eff}}$), Gaussian/tail DOS terms, and mobility prefactor ($\mu_0$)---not direct extractions of defect density, oxygen-vacancy concentration, microscopic mobility, or trap parameters~\cite{kamiya2010igzo}.

\begin{figure*}[!t]
\centering
\includegraphics[width=\textwidth]{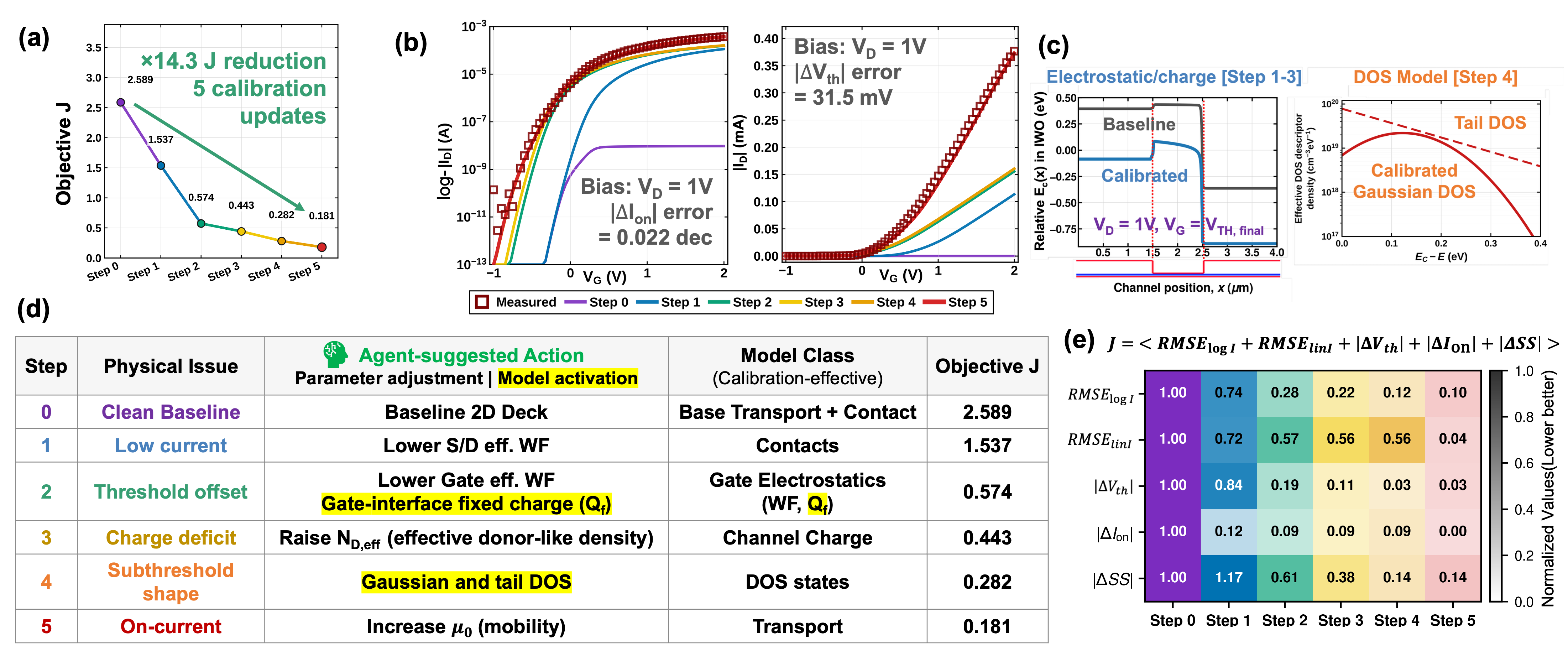}

\caption{Agent-suggested residual-to-physics updates yield a calibrated BG-IWO TCAD model. (a) Five accepted updates reduce $J$ from 2.589 to 0.181. (b) Stepwise transfer curves show progressive correction. (c) Calibrated electrostatic and DOS panels show how agentic actions address systematic residuals. (d) The update table links each residual component to the agent-suggested TCAD action and model class. Yellow highlights mark model activations, showing that $Q_f$ and DOS terms are introduced only after parameter adjustments leave systematic residual errors. (e) The residual heatmap shows which error components are reduced at each calibration step. Each value is normalized to its Step~0 value for visualization (Step~0 = 1).} 

\label{fig:trajectory}
\end{figure*}

\subsection{Fixed-Model Evaluation Across Bias and Geometry}
Fig.~6 supports generalization of the 2\%-W device model beyond the nominal calibration condition. The same fixed TCAD parameter set is used for the varying-drain-bias tests at \(V_D=\{0.05,0.10,0.20,0.50,1.00\}~\mathrm{V}\) and the varying-channel-length tests at \(L=\{1,2\}~\mu\mathrm{m}\) without retuning as described in Table~\ref{tab:data_splits}. The average and maximum errors are 24.9 and 36.1~mV, respectively, in $V_{\mathrm{th}}$, and 0.013 and 0.022 decade, respectively, in $I_{\mathrm{on}}$ for the drain-bias tests. For the channel-length tests, the corresponding average and maximum errors are 27.4 and 46.2~mV in $V_{\mathrm{th}}$ and 0.050 and 0.062 decade in $I_{\mathrm{on}}$. Because these splits are defined before calibration, the results demonstrate transferability across bias and geometry, rather than another local parameter fit.

\begin{figure*}[!t]
\centering
\includegraphics[width=\textwidth]{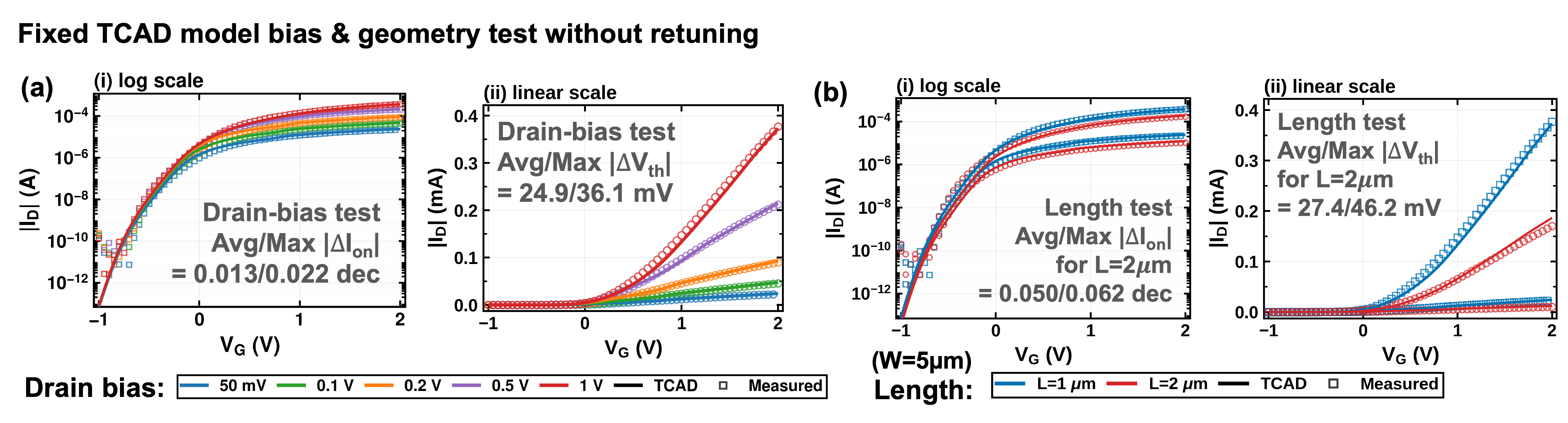}
\caption{Fixed TCAD model evaluation across drain bias and geometry. (a) Varying-drain-bias and (b) varying-channel-length tests of the fixed calibrated 2\%-W BG-IWO model. The same TCAD physics models and parameter set are used without retuning, demonstrating model transfer beyond the nominal calibration conditions.}
\label{fig:generalization}
\end{figure*}

\FloatBarrier

\section{Process-Sensitive Insight}
Fig.~7 uses the 2\%-W model as the reference and evaluates the separately recalibrated 4\%-W split for process-composition sensitivity. Fig.~7(a) shows that the 4\%-W device has a positive $V_{\mathrm{th}}$ shift and lower current relative to the 2\%-W reference. The observed W-composition trend is consistent with prior InO$_x$ studies showing that stronger dopant--oxygen bonding reduces sensitivity to oxygen processing and can suppress carrier-density increases associated with oxygen release~\cite{aikawa2013dopants}. In TCAD, this trend is represented by a reduced effective donor-like background density \(N_{D,\mathrm{eff}}\), rather than by microscopic extraction of oxygen-vacancy density. 

The agent-suggested sequence in Fig.~7(b) proceeded as follows. To address the dominant positive $V_{\mathrm{th}}$ shift and current reduction observed in the 4\%-W device relative to the 2\%-W reference, Step~1 reduced $N_{D,\mathrm{eff}}$ from $2.56\times10^{18}$ to $6.39\times10^{16}~\mathrm{cm}^{-3}$. To correct the residual $V_{\mathrm{th}}$ offset through gate electrostatics, Step~2 increased the gate WF from 4.00 to 4.45~eV and changed $Q_f$ from $+1.59\times10^{10}$ to $-1.00\times10^{12}~\mathrm{cm}^{-2}$. To reduce the remaining SS residual, Step~3 broadened the acceptor-tail distribution from $\sigma_{\mathrm{tail}}=0.133$ to 0.238~eV, increased the donor-like Gaussian DOS from $2.19\times10^{19}$ to $4.40\times10^{19}~\mathrm{cm}^{-3}\mathrm{eV}^{-1}$, shifted its center toward $E_C$ from $E_C-0.122$ to $E_C-0.080$~eV, and increased its width from $\sigma=0.080$ to 0.224~eV.

Fig.~7(c) shows the sequential TCAD $I_D$--$V_G$ progress as these descriptors move the curve toward the 4\%-W data. Here, the calibrated TCAD descriptor updates provide a model-level inverse diagnostic of W-composition sensitivity by identifying the model and parameter classes that reduce distinct components of the 4\%-W measurement--TCAD residual. The resulting sequence provides process-sensitive insight that can guide subsequent process experiments and TCAD studies.

\FloatBarrier

\begin{figure}[!tbp]
\centering
\figureasset{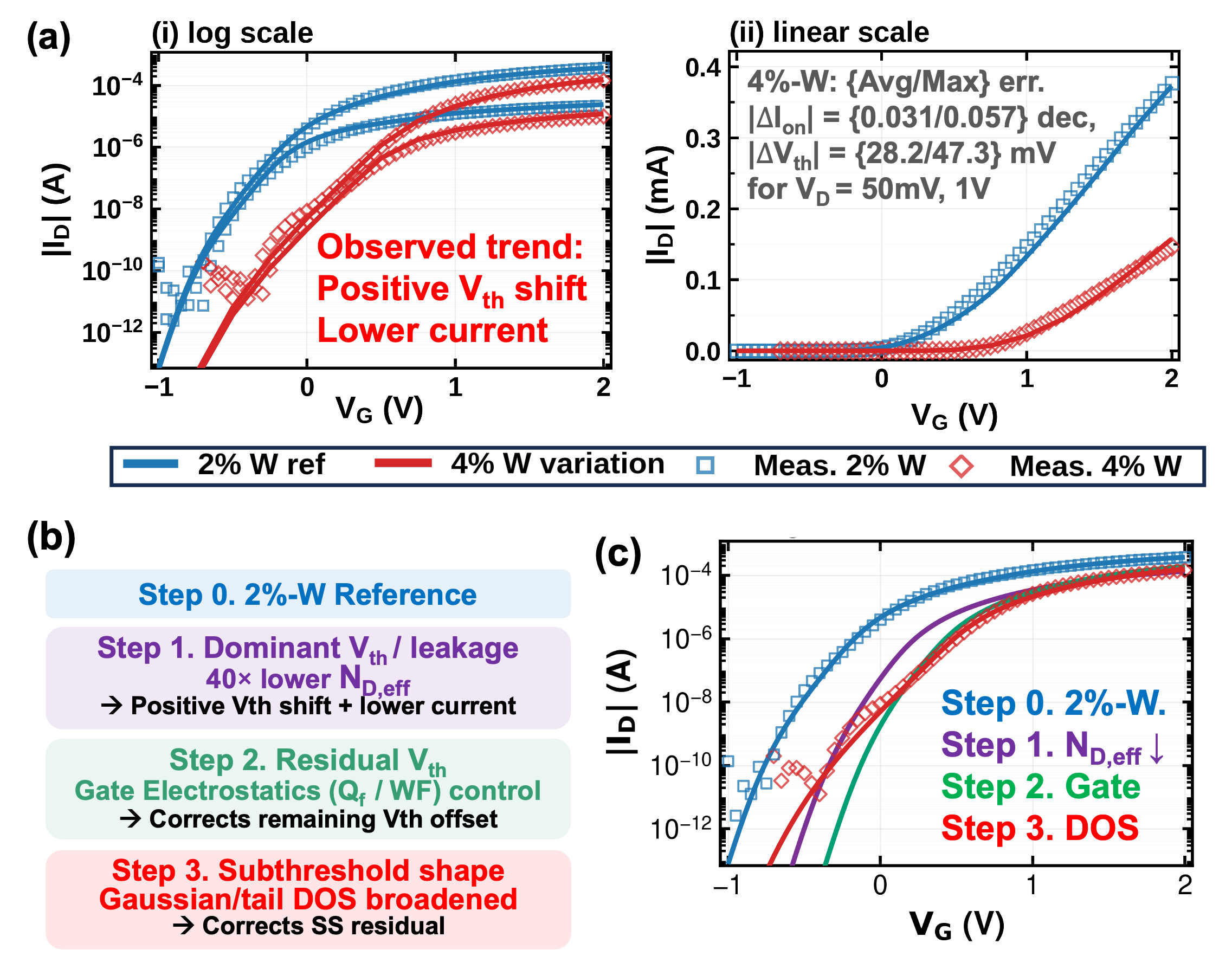}{}{}
\caption{W-composition recalibration test identifies process-sensitive insight. (a) $I_D$--$V_G$ characteristics of BG-IWO transistors with different W compositions: 2\% reference vs. 4\% variation. The 4\%-W device shows a positive $V_{\mathrm{th}}$ shift and lower current relative to the 2\%-W reference. The 4\%-W fitting and metrics use its measured range, $V_G \ge -0.7~\mathrm{V}$. (b) The agent-suggested descriptor sequence assigns the dominant $V_{\mathrm{th}}$ and current shifts to a lower effective donor-like background density $N_{D,\mathrm{eff}}$, the residual threshold offset to gate electrostatics, and the subthreshold shape to broadening of the Gaussian and tail DOS terms. (c) Sequential TCAD progress shows each descriptor moving the curve toward
the 4\%-W data ($V_D = 1~\mathrm{V}$).}
\label{fig:composition}
\end{figure}
\FloatBarrier

\section{Conclusion}
An agentic TCAD calibration workflow is demonstrated for fabricated BG-IWO oxide semiconductor transistors using measured $I_D$--$V_G$. The framework converts device context into a human-reviewed TCAD calibration setup and constrains deck updates through reviewed model classes and parameters. For the reference 2\%-W model, the workflow reduces the device objective \(J\) by 14.3\(\times\) through the residual-guided calibration loop. The fixed calibrated model passes drain-bias and channel-length transferability tests. The 4\%-W process-composition test provides process-sensitive insight by ranking the TCAD model classes that reduce distinct residual components associated with the W-composition change. This agentic TCAD calibration framework uses an LLM agent that orchestrates the workflow, while validity is established by the reviewed model classes, live Sentaurus execution, and transferability tests. The proposed methodology is extendable to other emerging device technology pathfinding to allow a fast iterative loop between model generation and process optimization.

\section*{Acknowledgment}
This work was sponsored by Samsung Electronics (IO250304-12193-01) and by in-kind support from NVIDIA.

AI-use disclosure: The authors used generative AI to assist the agentic TCAD workflow study. All the reported results were reviewed and verified by the authors.

\end{document}